\documentclass[final,1p,times]{elsarticle} 
\usepackage{graphicx} 
\usepackage{amssymb} 
\usepackage{amsthm} 
\usepackage{lineno} 

\journal{Nuclear Physics A} 
\begin{document} 

\begin{frontmatter} 


\title{Quarkyonic Matter and the Revised Phase Diagram of QCD}

\author{Larry McLerran$^{a}$ }

\address[a]{RIKEN Brookhaven Center and Brookhaven National Lab., 
Physics Dept.,
Upton, NY 11973-5000 USA}

\begin{abstract} 
At high baryon number density, it has been proposed that a new phase of QCD matter
controls the physics.  This matter is confining but can have densities much larger
than $\Lambda_{QCD}^3$.  Its existence is argued from large $N_c$ approximations,
and model computations.  It is approximately chirally symmetric.
\end{abstract} 

\end{frontmatter} 



\section{Introduction}

\begin{figure}[ht]
\centering
\includegraphics[scale=0.30]{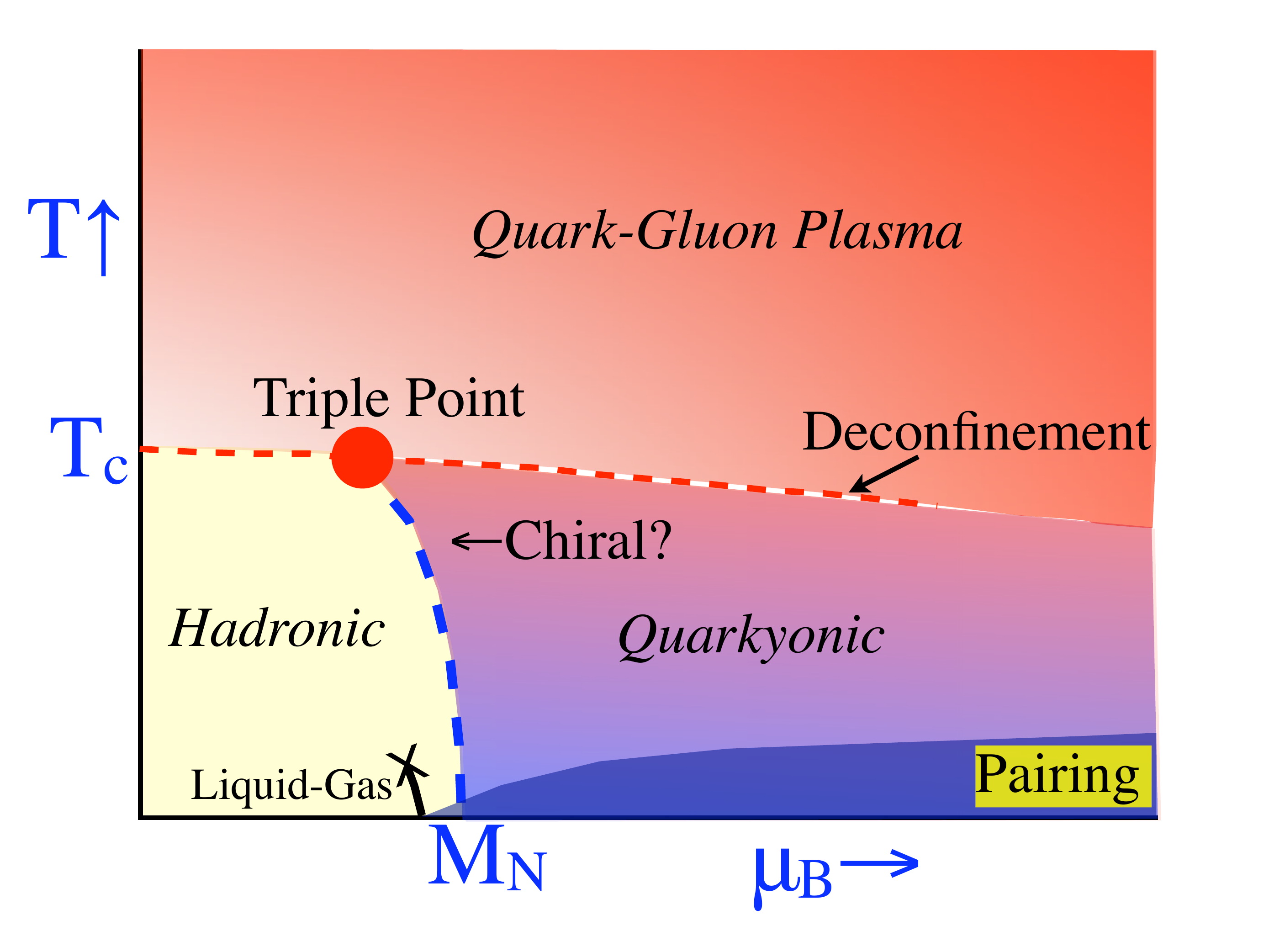}	       
\caption[]{The revised phase diagram of QCD}
\label{phasediagram}
\end{figure}

Rob Pisarski and I recently proposed the existence of Quarkyonic Matter \cite{McLerran:2007qj}.
These arguments were originally in the context of the large number of colors approximation,
$N_c \rightarrow \infty$ with the number of quark flavors held fixed, $N_f$ finite.  The arguments were then generalized to the large $N_c$ limit, with $N_f/N_c$ fixed  \cite{Hidaka:2008yy}.  Computations based on he PNJL model,\cite{Fukushima:2003fw},\cite{Ratti:2005jh},\cite{Pisarski:2000eq} were subsequently performed that argued that such a phase of matter may occur in systems with $N_c = 3$,
and that the quarkyonic transition {\it might} be associated with chiral symmetry restoration\cite{Fukushima:2008wg}-\cite{McLerran:2008ua}.  
In the paper with Redlich and Sasaki, a model was proposed that could continuously vary $N_c$,
and the behaviour of the phase transition boundaries could be studied as a function of $N_c$.\cite{McLerran:2008ua} A picture of the revised phase diagram of QCD is shown in Fig. \ref{phasediagram}  The purpose of this talk is to provide a brief explanation of these developments.

The work on Quarkyonic matter provides perhaps the most recent study of new phases of matter at high baryon number density.  There have been numerous speculations concerning high baryon number density matter, the most recent being work on color 
superconductivity \cite{Alford:1997zt},\cite{Rapp:1997zu}.
The properties of matter at finite baryon number density were in fact the first studies done in QCD concerning the properties of matter at very high energy density \cite{Collins:1974ky},
and a history of the early developments in this area are reviewed in Ref. \cite{McLerran:2008qi}.
An issue that the work on Quarkyonic matter seeks to clarify is the relationship between confinement
and high energy density matter.  Our conclusion is somewhat surprising:  Quarkyonic matter can
have densities parametrically large compared to the QCD scale and still be confined.  The experimental study of such matter motivated early discussion of ultra-relativistic heavy ion collisions  \cite{Stoecker:1980uk}, \cite{Anishetty:1980zp}, and will be the subject of future studies at RHIC, FAIR and NICA.

\section{The Large $N_c$ Approximation}

The large $N_c$ approximation is the limit where the number of colors of $QCD$ is taken to be very large, but the coupling strength $\lambda = g^2 N_c/4\pi$ is fixed.  This theory is asymptotically free, and 
confining in vacuum.  There is a Hagedorn spectrum of mesons, and mesons made of quarks interact with strength $1/N_c$ and glueballs with strength $1/N_c^2$.  Baryons are made of $N_c$ quarks,
have a mass of order $N_c\Lambda_{QCD}$ and interact strongly with strength $N_c$.

At finite temperature and zero baryon number density, the low temperature phase is composed
of light mass mesons.  The number of degrees of freedom are of order one in the number of colors.
At very low temperature, the Goldstone degrees of freedom are important, and there are $N_f^2-1$ degrees of freedom.  Baryon degrees of freedom vanish in the confined phase since
$e^{-M_B/T} \sim e^{-N_c}$
At very high temperatures, the degrees of freedom are
$2(N_c^2-1) + 4N_cN_f$ arising from the gluons and quarks.  Baryon degrees of freedom are unsuppressed in the Quark Gluon Plasma.

Since the number of degrees of freedom is of order one in the hadron phase and is of order $N_c^2$
in the Quark Gluon Plasma, there must be a jump in the energy density at a de-confinement temperature.  Viewed from the perspective of the hadron phase, thee is a limiting temperature where the energy density diverges.  For finite but large $N_c$, at the Hagedron limiting temperature, the density of
mesons becomes enormous, and at some point they must interact with one another.  As this occurs,
one makes a transition to a Quark Gluon Plasma, that has a large but finite number of degrees of freedom.  The rapid cross over for QCD at $N_c = 3$ is a remnant of this Hagedorn transition of large $N_c$ QCD \cite{Thorn:1980iv}.

At finite baryon density, we see that $e^{(\mu_B-M_B)/T}$ will be of order $e^{-N_c}$ for
$\mu_B$ much less than $M_b$.  In fact unless $( \mu_B-M_N)/M_N \sim 1/N_c$, the baryon number density is exponentially small.  Therefore when $\mu_B$ is $M_N$, there is a baryon number generating transition.
Of course, due to the very strong interactions of baryons in the large $N_c$ limit, the threshold
for baryon number generation is not the free nucleon mass, but is the mass of a nucleon in bound
nuclear matter.  

High density baryonic matter however remains confined.  The contribution of fermion loops
is suppressed by $\lambda \mu_Q^2/N_c$, where $\mu_Q = \mu_B/N_c$ is a quark chemical potential.
Such loops cannot Debye screen the confining potential until $\mu_Q \sim \sqrt{N_c} \Lambda_{QCD}$
This means that the de-confinement phase transition is independent of baryon density
In spite of the fact that the baryonic matter is confined, at high density the typical bulk quantities such
as pressure and energy density should be described by a weakly coupled gas of quarks.  This is because for particles inside a Fermi sea, interactions are not infrared sensitive, and bulk quantities should be dominated by quark kinetic energies.

This matter that is simultaneously confined and yet has its energy density and pressure well approximated by a quasi-free gas of quarks, Rob Pisarski and I have named as Quarkyonic.

For $N_c = 3$, the phase transitions of the large $N_c$ world can become cross overs.  The Quarkyonic transition remains at $\mu_B = M_N$ for $T = 0$, since this is the physical threshold for producing baryon number.  As T is increased, presumably the transition is at small $\mu_B$.  At high density, the
confinement transition should become weaker since the effect of quarks will begin to reduce the effect of the confinement temperature at higher density.  It also should move to lower values of T as $\mu_B$ increases.  The width of the Quarkyonic transition is estimated from the baryon chemical potential which for Fermi momenta $k_F << M_N$,  $\mu_B = M_N + k_F^2/2M_N = M_N(1 + O(1/N_c^2)$,
since the transition should be completed when $k_F \sim \Lambda_{QCD}$.

\section{Physical Picture of Quarkyonic Matter}

The Hadron Phase is composed of mesons and glueballs.  In the large $N_c$ limit, there are no baryons.
The number of Goldstone degrees of freedom are $N_f^2-1$
The Quark Gluon Plasma is composed of quarks and gluons with $2(N_c^2-1) + 4 N_fN_c$ degrees of freedom.

The Quarkyonic phase is composed of quarks, mesons and glueballs.  The number of quark degrees of freedom are $2N_cN_f$ and in addition there are $N_f^2-1$ Goldstone boson degrees of freedom.
The transition between these various phases should be thought of as a change in the number of degrees of freedom in a narrow range of temperature and baryon chemical potential.  A phenomenological parameterization that embodies these constraints should be straightforward.

The quarkyonic phase can be thought of as a Fermi gas composed of quasi-free quarks.  Near the Fermi surface the degrees of freedom are confined baryons.  The thermal excitations are mesons and glueballs.  Color superconductivity in the quarkyonic phase would not be allowed in the large $N_c$ limit, but surely for finite $N_c$ such phenomena are possible. 

One can understand the Quarkyonic transition in large $N_c$ with fixed $N_c/N_f$ in a simple way:  For large $N_f$ there are exponentially
large numbers of degenerate lowest mass baryon states, $N \sim e^{N_cF(N_c/N_f)}$.  Near the Quarkyoninc transitions, these states become important, since the probability that they contribution can be of order $e^{(N_cF(N_c/N_f) \mu_B/T - M_N/T)}$  The accumulating density is however rapidly cutoff by the strong interactions of the baryons.

The nature of chiral symmetry breaking is not so simple to understand.  In the high density phase,
one expects that the effects of high density matter may lower or perhaps reduce to zero the chiral quark condensate \cite{Glozman:2008ja},\cite{Glozman:2008kn},\cite{Glozman:2008jw},\cite{Glozman:2008fk},\cite{Glozman:2009sa}.  Intuitively, this is because a high baryon density will exclude quarks unless they have energy near the Fermi surface.  This means that if pairs form, they arise from particle hole pairs near the Fermi surface .  It is indeed possible that pairing phenomena near the Fermi surface might generate chiral symmetry breaking through chiral density waves \cite{Shuster:1999tn}-\cite{Park:1999bz}.  These pairs would break the translational invariance and a crystal would result \cite{Klebanov:1985qi}.  (In previous considerations of chiral density waves, it was assumed that the potential between quarks was Coulombic.  In Quarkyonic matter, the potential should be linear, and the possibility of condensation may be easier to realize.)

One can question whether or not the large $N_c$ limit is applicable for $N_c = 3$.  On the plus side of the equation is the fact that quarks seem to little influence the confining potential in vacuum  out to a distance of about $2~fm$, as seen from lattice Monte-Carlo computations.  The linear potential begins at a distance of about $0.2~fm$.  One may expect that there is some finite range of density before media
quarks may short out the potential, so there is room for a Quarkyonic phase.  There have also been model computations that argue for a Quarkyonic phase \cite{Fukushima:2008wg},\cite{McLerran:2008ua}.  These computations also argue that the chiral transition is approximately coincident with the 
Quarkyonic transition.

The negative side follows from some generic features of the large $N_c$ limit that are not observed in baryons.  The large $N_c$ limit predicts that  pion exchange has an intrinsic strength of order $N_c$ at large distances.  This requires that the ground state of large $N_c$ baryonic matter is a Skyrme crystal \cite{Klebanov:1985qi}.  On the other hand nuclear matter for $N_c = 3$ is a liquid, not a solid,
and the biding energy is of order $15~MeV$, not of order the nucleon mass.  Moreover, lattice gauge theory computations argue that at long and intermediate distance, the nuclear force is not of strength of order $N_c$.  How these facts become consistent with the large $N_c$ approximation is a mystery.
Perhaps the small binding energy of nuclear matter is an accident peculiar to $N_c = 3$.  If not, then we do not understand something fundamental about the nature of QCD.

\section*{Acknowledgments}
I thank my colleague Rob Pisarski, Yoshimasa Hidaka, Krzysztof Redlich, Chhiro Sasaki and Toru Kojo with whom I have collaborated in the development of these ideas.
The research of  L. McLerran is supported under DOE Contract No. DE-AC02-98CH10886.

\end{document}